\newcounter{myctr}
\def\myitem{\refstepcounter{myctr}\bibfont\noindent\ifnum\themyctr>9\else\phantom{0}\fi\hangindent17pt\themyctr.\enskip}

\documentclass{ws-ijqi}
\usepackage{hyperref}
\usepackage[super,sort,compress]{cite}
\usepackage{url}
\usepackage{bbm}
\usepackage{epstopdf}
\usepackage{color}

\newcommand{\ket}[2][]{{|#2\rangle_{#1}}}
\newcommand{\bra}[2][]{{}_{#1}\langle #2|}

\newcommand{\proj}[2][]{\ket{#2}_{#1}\bra{#2}}

\renewcommand{\t}[1]{\textrm{#1}}
\newcommand{\openone}{\mathbbm 1}

\begin{document}

%%%%%%%%%%%%%%%%%%%%% Publisher's Area please ignore %%%%%%%%%%%%%%
\catchline{}{}{}{}{}
%%%%%%%%%%%%%%%%%%%%%%%%%%%%%%%%%%%%%%%%%%%%%%%%%%%%%%%%%%%%%%%%%%%

\title{ON SUPERRESOLUTION IMAGING AS A MULTIPARAMETER ESTIMATION PROBLEM}

\author{ANDRZEJ CHROSTOWSKI and RAFA{\L} DEMKOWICZ-DOBRZA\'{N}SKI}

\address{Faculty of Physics, University of Warsaw,\\ul.\ Pasteura 5, 02-093 Warszawa, Poland}

\author{MARCIN JARZYNA and KONRAD BANASZEK$^\ast$}

\address{Centre of New Technologies, University of Warsaw\\
ul.\ Banacha 2c, 02-097 Warszawa, Poland\\
$^\ast$k.banaszek@cent.uw.edu.pl}

\maketitle

\begin{history}
\received{Day Month Year}
\revised{Day Month Year}
%\accepted{Day Month Year}
%\comby{(xxxxxxxxxx)}
\end{history}

\begin{abstract}
We consider the problem of characterising the spatial extent of a composite light source using the superresolution imaging technique when the centroid of the source is not known precisely. We show that the essential features of this problem can be mapped onto a simple qubit model for joint estimation of a phase shift and a dephasing strength.
\end{abstract}


%\tableofcontents  % optional

\markboth{A. Chrostowski, R. Demkowicz-Dobrza\'{n}ski, M. Jarzyna \& K. Banaszek}
{On Superresolution Imaging as a Multiparameter Estimation Problem}

\section{Introduction}

The promise of superresolution imaging is to determine characteristics of a spatially extended light source with precision better than that defined by the diffraction limit \cite{TsangNairPRX2016,NairTsangPRL2016}. The diffraction limit is a consequence of a finite aperture of the optical instrument used for observation combined with the conventional measurement of the spatial distribution of light intensity in the image plane.\cite{BornWolfPrinciplesofOptics} At the fundamental level, an intensity measurement consists of registering a finite number of counts generated by incident photons. Generation of each photocount is an inherently random process that needs to be described in statistical terms.\cite{MandelWolfOpticalCoherence} The randomness of locations where individual photons are registered effectively masks features of a composite light source whose extent is below the diffraction limit. The reconstruction of such features by conventional means would require collection of an immense amount of data to suppress the effects of statistical uncertainty.\cite{BobroffRSI1986}

As proposed by Tsang and Nair\cite{TsangNairPRX2016,NairTsangPRL2016} and demonstrated in a series of proof-of-principle experiments,\cite{PaurStoklasaOPTICA2016,YangTashilinaOPTICA2016} the above difficulty can be overcome by detecting the incident photons after separating them in a carefully selected basis of spatial modes. This technique, called mode demultiplexing, requires additional {\em a priori} knowledge of the centroid of the light source. The most straightforward approach to deal with this issue is to perform a standard spatially resolved measurement on a fraction of available photons and to estimate the centroid as the average position.\cite{TsangNairPRX2016} In this contribution we consider the problem of simultaneous determination of the centroid and the spatial extent of a composite light source. We show that when these two parameters are well below the diffraction limit, the problem can be modelled with the help of an elementary qubit system in which the parameters of interest correspond respectively to a rotation and a contraction of the Bloch vector. This observation links supperresolution imaging to quantum multiparameter estimation which has been addressed in several recent works.\cite{PerezDelgadoPearcePRL2012,VidrighinDonatiNCOMM2014,RagyJarzynaPRA2016,PearceCampbellQuantum2017}

This paper is organised as follows. In Sec.~\ref{Sec:Demultiplexing} we review the spatial mode demultiplexing technique for the determination of the spatial extent of a composite light source when its characteristic size is well below the diffraction limit. The measurement error is discussed in Sec.~\ref{Sec:Error}. The qubit model for the underlying estimation problem is presented in Sec.~\ref{Sec:Qubitmodel}. Finally, Sec.~\ref{Sec:Conclusions} concludes the paper.

\section{Spatial mode demultiplexing}	%) A SECTION HEADING
\label{Sec:Demultiplexing}

\begin{figure}[t]
\includegraphics[scale=0.333]{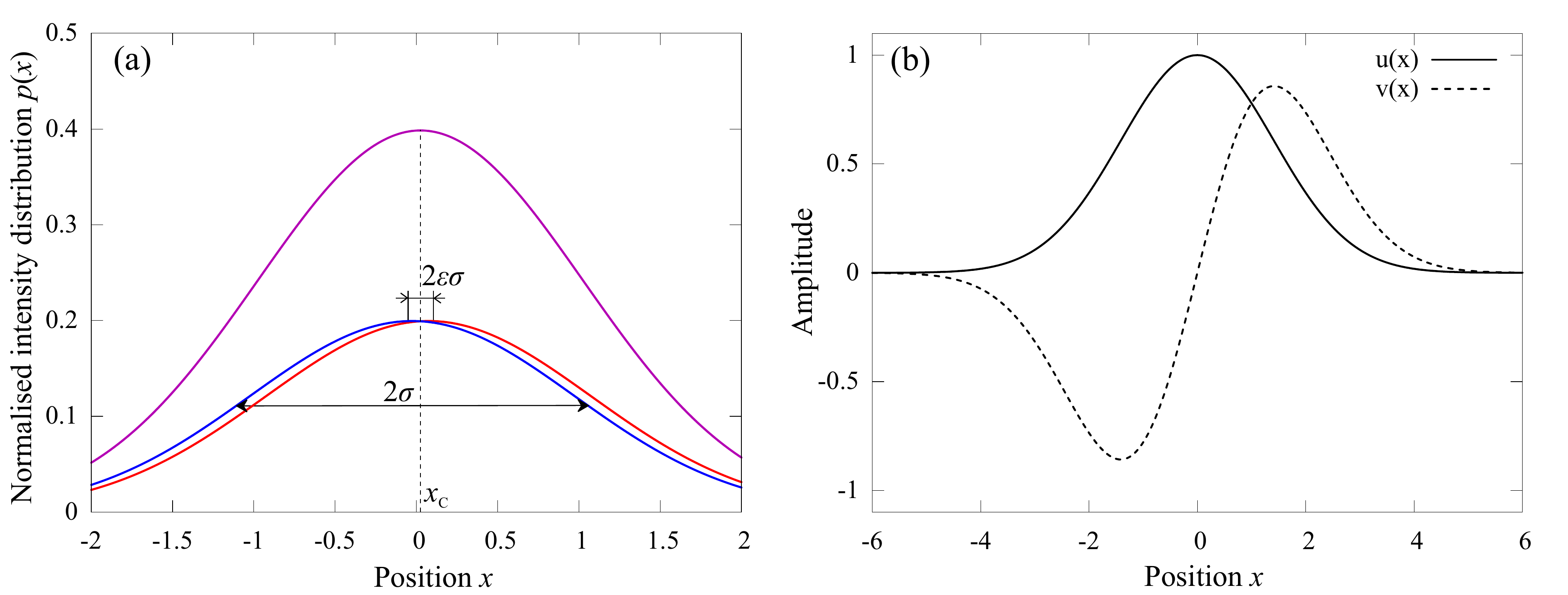}\centering
\caption{(a) Spatial intensity distribution produced by a pair of incoherent point sources separated by $2d=0.01 $ with centroid located at $x_C = 0.025 $ with respect to the reference position $x_R=0$. Each individual source produces a Gaussian distribution characterised by standard deviation $\sigma=1$. The problem of characterising the extended source can be described with two dimensionless parameters $\varepsilon = d/\sigma$ and $\theta = (x_C- x_R)/\sigma$. (b) The amplitude transfer function $u(x)$ and its normalised derivative $v(x)$ used in the spatial mode demultiplexing technique.}
\label{Fig:Principle}
\end{figure}

Consider an ensemble of mutually incoherent point sources labelled with an index $j$ characterised by relative strengths $w_j$, $\sum_j w_j = 1$.
For simplicity we will discuss image formation using one spatial dimension. We will also assume that the probability of detecting more than one photon in a given observation time interval is negligibly small and hence we will think of image formation as a series of repeated single photon detection events.
In the image plane each source generates a coherent field distribution described by an amplitude transfer function $u(x-x_j)$ with its centre located at $x_j$. We will assume that the transfer function $u(x)$ is real and even, i.e.\ $u(x) = u^\ast(x) = u(-x)$, and furthermore that its square is normalized to one, $\int_{-\infty}^{\infty} \t{d}x \,[u(x)]^2 = 1$.  The objective is to determine the spatial extent of the sources in a scenario when they are spread over a range much smaller than the characteristic width of the transfer function, corresponding to the diffraction limit. In the standard spatial intensity distribution measurement, the probability density for detecting a photon at a given point $x$ is  given by a weighted sum
\begin{equation}
p(x) = \sum_j w_j [u(x-x_j)]^2.
\label{Eq:p(x)}
\end{equation}
In the limit of small spreads, this distribution can hardly be distinguished from the one corresponding to a single point source located at the centroid of the source $x_C = \sum_{j}w_j x_j$, in which case it would read $p(x)=[u(x-x_C)]^2$. This is illustrated in Fig.~\ref{Fig:Principle}(a) with the example of two equally weighted point sources separated by a distance $2d$ much smaller than the width $\sigma$ of the transfer function, assumed to have the Gaussian form
\begin{equation}
u(x) = \frac{1}{\sqrt[4]{2\pi\sigma^2}} \exp\left( -\frac{x^2}{4\sigma^2} \right).
\label{Eq:u(x)Gaussian}
\end{equation}
This difficulty leads to an intuitive expectation---the Rayleigh's criterion---that estimating the separation between the sources in this regime is subject to large uncertainty. Using methods of parameter estimation theory that provide general quantitative bounds on how precisely a parameter can be estimated within a given probabilistic model, this intuition can be formulated in a rigorous way confirming that in the case of a direct measurement of the spatial intensity distribution, the precision indeed deteriorates significantly in the small separation regime irrespectively of what inference strategy one pursues. \cite{Bettens1999,TsangNairPRX2016, Tsang2016}. In particular, for the given example of two sources separated by distance $2d$ with the Gaussian transfer function (\ref{Eq:u(x)Gaussian}), the precision behaves as $\Delta d \approx \frac{\sigma^2}{d}\sqrt{\frac{2}{N}}$, where $N$ is the number of registered photons. This formula clearly shows divergence in the limit of small separations $d\to 0$.

The basic idea of superresolution imaging based on mode demultiplexing is to measure the intensity of incoming radiation in a basis of spatial modes that provides a signal more sensitive to the extent of the ensemble. In the simplest model valid for small extents, it is sufficient to consider a mode function proportional to the derivative of the transfer function
\begin{equation}
v(x) = - 2 \sigma \frac{\t{d} u}{\t{d} x}.
\label{Eq:vdef}
\end{equation}
The multiplicative factor warrants the normalisation of $v(x)$. To satisfy this condition for a general transfer function that is not necessarily Gaussian, the parameter $\sigma$ should be taken as
\begin{equation}
\sigma = \frac{1}{2} \left( \int_{-\infty}^{\infty} \t{d} x \left( \frac{\t{d} u}{\t{d} x} \right)^2\right)^{-1/2}.
\end{equation}
For a real transfer function $u(x)$ the functions $u(x)$ and $v(x)$ are mutually orthogonal, i.e.\
$\int_{-\infty}^{\infty} \t{d} x \, v(x) u(x)=0$. An exemplary set of mode functions $u(x)$ and $v(x)$, assuming a Gaussian shape for the former given explicitly by
Eq.~(\ref{Eq:u(x)Gaussian}) is shown in Fig.~\ref{Fig:Principle}(b). The functions are analogous to the ground state and the first excited state of a quantum mechanical harmonic oscillator in the position representation. The modes $u(x)$ and $v(x)$ can be separated using integrated optics structures\cite{TsangNairPRX2016} or a spatial light modulator.\cite{PaurStoklasaOPTICA2016}

Suppose now that instead of conventional spatially resolved detection, the incoming light is demultiplexed in the basis of spatial modes $u(x-x_R)$ and $v(x-x_R)$ centred at a reference point $x_R$ and that the intensity of individual components is measured. The probability that an incoming photon is detected in the mode $v(x)$ is given by
\begin{equation}
I(x_R) = \sum_j w_j \left( \int_{-\infty}^{\infty} \t{d} x \, v(x-x_R) u(x-x_j) \right)^2 \approx \frac{1}{(2\sigma)^2} \sum_j w_j (x_j - x_R)^2,
\label{Eq:Intensityv}
\end{equation}
where in the second step we have expanded $u(x-x_j) \approx u(x-x_R) + \frac{1}{2\sigma} (x_j-x_R) v(x-x_R)$ up to the linear term and the orthogonality of the mode functions $u(x)$ and $v(x)$ has been used. Note that $I(x_R)$ is the fraction of the source intensity directed to the mode $v(x-x_R)$.
Intrinsic properties of the composite source can be determined if the above measurement is performed at the centroid of the system $x_C$, given by the weighted sum $x_C = \sum_j w_j x_j$, in which case
\begin{equation}
 I(x_C) = I_C = \frac{1}{(2\sigma)^2} \sum_{j} w_j d_j^2,
 \label{Eq:IC}
\end{equation}
where $d_j = x_j - x_C$ denote relative distances of individual sources from the system centroid and satisfy $\sum_{j} w_j d_j = 0$.
The right hand side of the above formula has a simple interpretation as the second moment for the distribution of point sources with respect to the centroid of the system, expressed in units $(2\sigma)^2$.

\section{Measurement error}
\label{Sec:Error}

As a concrete example, suppose that the system comprises two point sources of equal brightness separated by the distance $2d$, which implies that  $I_C = d^2/(2\sigma)^2$.
Let us consider a situation where there are $N$ incoming photons in total. If the centroid is known perfectly, they are detected in the mode basis $u(x-x_C)$, $v(x-x_C)$ with respective probabilities $p_{u} \approx 1-I_C$, $p_{v} \approx I_C$. A general result
on the asymptotic efficiency of maximum likelihood estimation \cite{kay1993fundamentals} implies that
in the limit of large $N$ the optimal precision of estimating $d$
is given by
\begin{equation}
\Delta d = \frac{1}{\sqrt{N F}},\quad  F = \sum_{i=u,v} \frac{1}{p_i} \left(\frac{\partial p_i}{\partial d}\right)^2,
\end{equation}
where $F$ is the Fisher information. In our case $F = [\sigma^{2}(1-\frac{d^2}{4\sigma^2})]^{-1}$, and hence the
parameter $d$ can be estimated with uncertainty equal to:\cite{TsangNairPRX2016}
\begin{equation}
\label{eq:separation}
\Delta d= \frac{\sigma}{\sqrt{N}}\left(1-\frac{d^2}{4\sigma^2}\right)^{1/2}.
\end{equation}
This provides a huge advantage over the spatial intensity distribution measurement and allows to circumvent the Rayleigh criterion since the estimation uncertainty
does not diverge even when taking the limit  ${d}/{\sigma} \rightarrow 0$. Note that the above derivation is valid only in the regime $d/\sigma \ll 1$.

What happens if one does not know exactly the location of the centroid?
If the measurement is performed with respect to a general position $x_R$ a straightforward calculation using the approximate expression derived in Eq.~(\ref{Eq:Intensityv}) yields the probability of detection equal to
\begin{equation}
I(x_R) = I_C + \frac{1}{(2\sigma)^2} (x_R - x_C)^2.
\label{Eq:Steiner}
\end{equation}
    Note that this formula is analogous to Steiner's theorem for the mass moment of inertia. The second term in Eq.~(\ref{Eq:Steiner}), originating from the imperfect knowledge of the centroid, constitutes a systematic error. In the standard approach, the location $x_R$ is determined via spatially resolved detection of a certain number photons characterized by the position distribution given in Eq.~(\ref{Eq:p(x)}). The variance of this distribution is
\begin{equation}
\label{eq:var}
\text{Var}(x) = \int_{-\infty}^{\infty} \t{d} x \, (x-x_C)^2 p(x) = \sum_j w_j d_j^2 + \int_{-\infty}^{\infty} \t{d}x \, x^2 [u(x)]^2 . %%%\approx \sigma^2,
\end{equation}
In the regime when the spatial extent of the source is well below the diffraction limit, the second term in the above expression, equal to the second moment of the squared transfer function, dominates the variance. Therefore for the Gaussian model considered here we have $\text{Var}(x) \approx \sigma^2$. By sacrificing $n$ photons to perform estimation of the centroid we may estimate its position with precision $\sigma/\sqrt{n}$. For $n$ large enough, more precisely $n^2 \gg I_C$, we may therefore make the second term in Eq.~(\ref{Eq:Steiner}) arbitrary small compared to the first one and estimate the parameter $d$ as before.
More formally, having the total of $N$ photons at our disposal, we can sacrifice $n \propto N^{\alpha}$, $0 < \alpha < 1$ to estimate the centroid, and keep the rest $N - N^{\alpha}$ for separation estimation. In the asymptotic limit of $N \rightarrow \infty$,
since $N - N^\alpha \approx N$, we should recover precision scaling as given in \eqref{eq:separation}. For any finite $N$ one needs to resort to numerical means to find the optimal partition of the photons used to measure the centroid and the extent of the source.

\begin{figure}[t]
\includegraphics[scale=0.333]{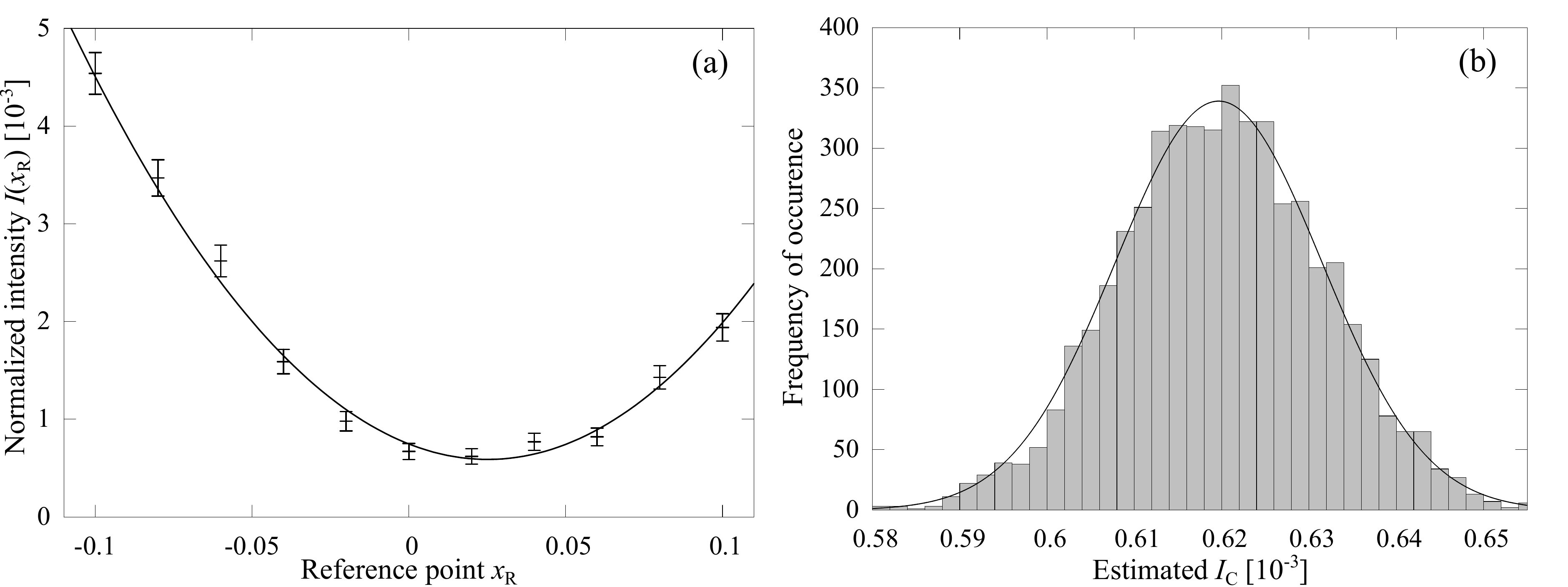}\centering
\caption{(a) Simulation of the measurement of the intensity $I(x_R)$ in the mode $v(x-x_R)$ with respect to a reference point $x_R$ assuming $N=10^5$ photons received from the source (crosses with vertical error bars) and the fitted parabolic function with the obtained values of free parameters $I_C$ and $x_C$. Actual parameters of the source are the same as those used in Fig.~\ref{Fig:Principle}. (b) A histogram of the values of $I_C$ obtained from 20000 repetitions of a simulation presented in the panel (a). The solid line is a Gaussian fit with the mean $0.6197(1) \times 10^{-3}$.}
\label{Fig:Simulation}
\end{figure}

An elementary method to deal with the problem of biasedness in determining $I_C$ would be to perform measurements at several locations $x_R$ in the vicinity of the centroid and then to fit a parabolic curve described by the right hand side of Eq.~(\ref{Eq:Steiner}) with $I_C$ and $x_C$ taken as free parameters.
An exemplary numerical simulation of this procedure with the same parameters as those used in Fig.~\ref{Fig:Principle}(a) is presented in Fig.~\ref{Fig:Simulation}(a).
In order to evaluate the accuracy of the procedure, we have performed $20000$ repetitions of the numerical simulation, which yielded the value $I_C = 0.6197(1) \times 10^{-3}$ as the average result. This is lower than $0.625 \times 10^{-3}$ calculated using Eq.~(\ref{Eq:IC}), but for a fair comparison one should use the exact integral expression given in Eq.~(\ref{Eq:Intensityv}), which yields the figure
$0.6246 \times 10^{-3}$ that remains higher than $I_C$ obtained from numerical simulations. Thus the quadratic fit method seems to produce an estimate that on average is slightly biased below the actual value of $I_C$
and in particular would underestimate the separation parameter $d$ for a pair of point sources.

\section{Qubit model}
\label{Sec:Qubitmodel}

\begin{figure}
\includegraphics[scale=0.5]{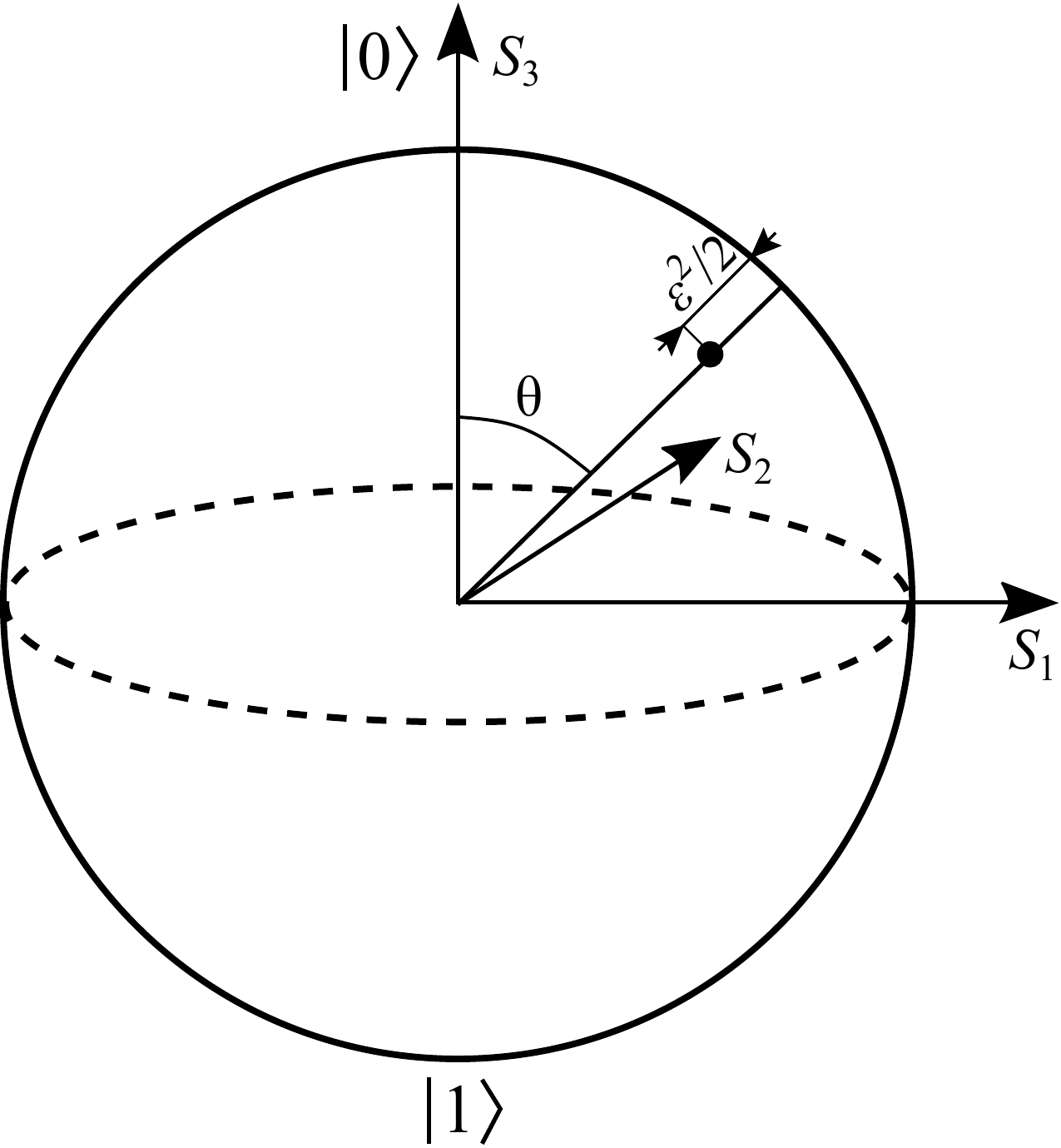}\centering
\caption{Bloch sphere representation of the qubit model for superresolution imaging at an uncertain location. The phase shift $\theta$ corresponds to the location of the source centroid with respect to the reference system and the reduction of the Bloch vector length by $\varepsilon^2/2$ can be related to the spatial extent od the source.}
\label{Fig:QubitModel}
\end{figure}

The essential features of the problem discussed in the preceding section can be distilled by considering a two-dimensional subspace spanned by the mode functions $u(x-x_R)$ and $v(x-x_R)$ that effectively defines a qubit. In order to keep the formulas concise, we will now switch to Dirac notation and use the following kets
\begin{equation}
u(x-x_R) \equiv \ket{0} , \qquad v(x-x_R) \equiv \ket{1}.
\end{equation}
The field from a source at location $x_j$ can be now written as:
\begin{equation}
u(x-x_j) \approx u(x-x_R) + \frac{1}{2\sigma} (x_j-x_R) v(x-x_R)  \equiv  \ket{0} + \frac{1}{2\sigma} (x_j-x_R) \ket{1}.
\label{Eq:u(x-x_j)qubit}
\end{equation}
Because the sources are mutually incoherent, the radiation needs to be described by the first-order coherence function\cite{MandelWolfOpticalCoherence} which in the Dirac notation takes the form of an operator:
\begin{multline}
\sum_j w_j u(x-x_j)u^\ast (x'-x_j) \\ %%%% \equiv \sum_j w_j \hat{D}(x_j-x_R)\proj{u}\hat{D}^\dagger(x_j-x_R) \\
\cong \proj{0} + \frac{1}{2\sigma}(x_C-x_R) \bigl( \ket{1}\bra{0} + \ket{0} \bra{1} \bigr)
+ \frac{1}{(2\sigma)^2} \left( (x_C-x_R)^2 +  \sum_j w_j d_j^2 \right)\proj{1}
\end{multline}
It will be convenient to introduce two dimensionless parameters that express relevant lengths in the units of $\sigma$: the location of the centroid
$\theta = (x_C-x_R)/\sigma$ and the effective radius of the source $\varepsilon = \left( \sum_j w_j (d_j/\sigma)^2 \right)^{1/2}$. The normalised density matrix written in the basis $\ket{0}, \ket{1}$ takes the form
\begin{equation}
\hat{\varrho}  = \frac{1}{1 + \frac{\varepsilon^2}{4} + \frac{\theta^2}{4}} \begin{pmatrix} 1 & \frac{\theta}{2} \\ \frac{\theta}{2} & \frac{\varepsilon^2}{4} + \frac{\theta^2}{4}  \end{pmatrix} .
\end{equation}
The role of the parameters $\varepsilon$ and $\theta$ can be most easily understood by considering the Bloch representation of the qubit state, $\hat{\varrho} = \frac{1}{2}\left(\openone + \sum_{i=1}^3 s_i \hat{\sigma}_i \right)$, where $\hat{\sigma}$ denote Pauli matrices.
The three components of the Bloch vector are given by
\begin{align}
s_1 & =  \frac{\theta}{1 + \frac{\varepsilon^2}{4} + \frac{\theta^2}{4}} \approx \left(1-\frac{\varepsilon^2}{2}\right) \sin \theta\nonumber \\
s_2 & =  0  \\
s_3 & = \frac{1 - \frac{\varepsilon^2}{4} - \frac{\theta^2}{4}}{1 + \frac{\varepsilon^2}{4} + \frac{\theta^2}{4}}  \approx \left(1-\frac{\varepsilon^2}{2}\right) \cos \theta . \nonumber
\end{align}
The second approximate expressions are correct up to the quadratic order in $\varepsilon$ and $\theta$. They offer a simple interpretation of our problem illustrated with Fig.~\ref{Fig:QubitModel}. In the great circle located in the plane $s_1, s_3$ the parameter $\theta$ corresponds to a rotation of the Bloch sphere about the axis $s_2$, whereas $\varepsilon$ is responsible for the contraction of the Bloch vector. This signals a connection of the studied problem with joint estimation of a phase shift and a dephasing strength.\cite{VidrighinDonatiNCOMM2014}

Using the introduced model, we may now
address the problem of estimating the parameters $\varepsilon$ and $\theta$  without
specifying a priori any particular measurement. Instead, we will calculate the quantum Fisher information (QFI) \cite{helstrom1976quantum} on the state $\hat{\varrho}$, which for a single parameter estimation problem equals the value of Fisher information
corresponding to the most informative measurement performed on the state. Specifically, the
QFI for the estimation of  a single parameter $\varepsilon$ can be calculated using the following formula:
\begin{equation}
F_Q = \t{Tr}(\hat{\varrho} \hat{L}_{\varepsilon}^2),
\end{equation}
 where $\hat{L}_{\varepsilon}$, called the symmetric logarithmic derivative, is given implicitly by $\partial_{\varepsilon} \hat{\varrho} = (\hat{L}_{\varepsilon} \hat{\varrho} + \hat{\varrho} \hat{L}_{\varepsilon})/2$.  As a result, the fundamental precision bound on estimating parameter $\varepsilon$, irrespectively of what measurement was performed, reads  $\Delta \varepsilon \geq 1/\sqrt{N F_Q}$, where $N$ is the number of repetitions of the experiment. In case of multiparameter estimation, one needs to define the QFI matrix, $(F_Q)_{\mu\nu} = \t{Tr}(\hat{\varrho} \hat{L}_\mu \hat{L}_\nu)$,
 where $\hat{L}_\mu$ is the symmetric logarithmic derivative corresponding to the parameter $\mu$.
 The inverse of the QFI matrix multiplied by the number of repetitions $(NF_Q)^{-1}$ provides a lower bound on the covariance matrix of the estimated parameters.
%%With the QFI matrix at hand, obtains a bound on the covariance matrix of estimated parameters in the form of the following matrix inequality  $\t{Cov} \geq (F_Q)^{-1}$.

The QFI matrix corresponding to the two-parameter estimation problem reads:
\begin{equation}
F_Q = \begin{pmatrix}
(F_Q)_{\varepsilon\varepsilon} & (F_Q)_{\varepsilon\theta}\\
(F_Q)_{\theta\varepsilon} & (F_Q)_{\theta\theta} \end{pmatrix}
=
 \begin{pmatrix}
1 + \frac{\varepsilon^2}{4} & 0  \\
0 & 1 - \varepsilon^2
\end{pmatrix},
\end{equation}
where the elements are specified up to the second order in the parameters $\varepsilon$ and $\theta$.
As a result, we obtain the following  bounds on the estimation precision in the leading order of $\varepsilon$:
\begin{equation}
\Delta \varepsilon \geq \frac{1}{\sqrt{N}}\left( 1- \frac{\varepsilon^2}{4} \right)^{1/2}, \qquad \Delta \theta \geq \frac{(1 + \varepsilon^2)^{1/2}}{\sqrt{N}}.
\end{equation}
Recalling the relation between $\varepsilon, \theta$ and the dimensional parameters $d, x_C-x_R, \sigma$,  we see that the bound on $\Delta \varepsilon$
corresponds exactly to the uncertainty formula for $d$ given in \eqref{eq:separation}, while the bound $\Delta \theta$
corresponds exactly to the estimation precision of the centroid as discussed below equation \eqref{eq:var}, when the first term
 in \eqref{eq:var} equal to $\sigma^2 \varepsilon^2$ is not neglected.
This shows that the measurements considered in Sec.~\ref{Sec:Error} are indeed optimal for the determination of one of the parameters $\varepsilon$ or $\theta$.
Unfortunately, these measurements are not compatible and cannot be performed jointly. This can also be seen explicitly in the qubit model considered here.
In general, the optimal measurements that maximize Fisher information with respect to a given parameter
are projection measurements in the eigenbasis of the respective symmetric logarithmic derivative operator.
At the operating point $\varepsilon = \theta = 0$ the corresponding eigenbases for $\hat{L}_{\varepsilon}$ and  $\hat{L}_\theta$ are $\ket{0},\ket{1}$ and $\ket{\pm} = (\ket{0} \pm \ket{1})/\sqrt{2}$ respectively. It is seen that these two measurements do not commute with each other. Hence, if one insists on using the optimal projective measurements, they need to be performed separately on different subsets of the input ensemble as discussed in Sec.~\ref{Sec:Error}. It is worth mentioning that while the symmetric logarithmic derivatives do not commute, their commutator yields zero when traced over the state $\hat{\varrho}$. This implies that the simultaneous
measurement reaching the optimal values of precision is possible provided collective measurements on many probes are performed \cite{RagyJarzynaPRA2016}.

\section{Conclusions}
\label{Sec:Conclusions}

We have presented an elementary discussion of the superresolution imaging technique based on spatial mode demultiplexing. In the regime when the spatial extent of a composite light source is much less than the diffraction limit, the demultiplexing method enables one to determine the second moment of the distribution of the constituent point sources. Intrinsic properties of the source are given by this second moment with respect to the centroid. If the exact position of the centroid is not known, it is necessary to adopt a multiparameter estimation approach. We have described a simple quantum mechanical model in which the spatial extent and the centroid of the source are analogues of the phase shift and the dephasing strength of a qubit. Interestingly, optimal projective measurements for estimating individually these parameters are mutually incompatible. The qubit model can provide insights in other imaging-related scenarios, e.g.\ involving hypothesis testing.\cite{arXiv:1609.00684}

\section*{Acknowledgments}
We acknowledge insightful disscussions with Saikat Guha and Wojciech Wasilewski.
This work is part of the project ``Quantum Optical Communication
Systems'' carried out within the TEAM programme of the Foundation for Polish Science cofinanced
by the European Union under the European Regional Development Fund. R.D.-D.\ acknowledges support of
National Science Center (Poland) grant No. 2016/22/E/ST2/00559.

\bibliographystyle{ws-ijqi}

\begin{thebibliography}{10}

\bibitem{TsangNairPRX2016}
M.~Tsang, R.~Nair and X.-M. Lu, {\em Physical Review X} {\bf 6}  (2016) p.
  031033.

\bibitem{NairTsangPRL2016}
R.~Nair and M.~Tsang, {\em Physical Review Letters} {\bf 117}  (2016) p.
  190801.

\bibitem{BornWolfPrinciplesofOptics}
M.~Born and E.~Wolf, {\em Principles of Optics: Electromagnetic Theory of
  Propagation, Interference and Diffraction of Light} (Cambridge University
  Press, Cambridge, 1999).

\bibitem{MandelWolfOpticalCoherence}
L.~Mandel and E.~Wolf, {\em Optical Coherence and Quantum Optics} (Cambridge
  University Press, Cambridge, 1995).

\bibitem{BobroffRSI1986}
N.~Bobroff, {\em Review of Scientific Instruments} {\bf 57}  (1986) 1152.

\bibitem{PaurStoklasaOPTICA2016}
M.~Pa{\'u}r, B.~Stoklasa, Z.~Hradil, L.~L. S{\'a}nchez-Soto and J.~Rehacek,
  {\em Optica} {\bf 3}  (2016) 1144.

\bibitem{YangTashilinaOPTICA2016}
F.~Yang, A.~Tashchilina, E.~S. Moiseev, C.~Simon and A.~I. Lvovsky, {\em
  Optica} {\bf 3}  (2016) 1148.

\bibitem{PerezDelgadoPearcePRL2012}
C.~A.~P\'{e}rez-Delgado, M.~E.~Pearce and P.~Kok, {\em Physical Review Letters}
{\bf 109} (2012) 123601.

\bibitem{VidrighinDonatiNCOMM2014}
M.~D. Vidrighin, G.~Donati, M.~G. Genoni, X.-M. Jin, W.~S. Kolthammer, M.~Kim,
  A.~Datta, M.~Barbieri and I.~A. Walmsley, {\em Nature Communications} {\bf 5}
   (2014) 3532.

\bibitem{RagyJarzynaPRA2016}
S.~Ragy, M.~Jarzyna and R.~Demkowicz-Dobrza{\'n}ski, {\em Physical Review A}
  {\bf 94}  (2016) 052108.

\bibitem{PearceCampbellQuantum2017}
M.~E.~Pearce, E,~T.~Campbell and P.~Kok, {\em Quantum} {\bf 1} (2017) 21.

\bibitem{Bettens1999}
E.~Bettens, D.~V. Dyck, A.~den Dekker, J.~Sijbers and A.~van~den Bos, {\em
  Ultramicroscopy} {\bf 77}  (1999) 37.

\bibitem{Tsang2016}
M.~{Tsang}, {\em arXiv preprint arXiv:1605.03799}  (May 2016)

\bibitem{kay1993fundamentals}
S.~M. Kay, {\em Fundamentals of Statistical Signal Processing: Estimation
  Theory} (Prentice-Hall, Upper Saddle River, 1993).

\bibitem{helstrom1976quantum}
C.~W. Helstrom, {\em Quantum Detection and Estimation Theory} (Academic Press,
  New York, 1976).

\bibitem{arXiv:1609.00684}
H.~Krovi, S.~Guha and J.~H. Shapiro, {\em arXiv preprint arXiv:1609.00684}
  (2016).

\end{thebibliography}

\end{document}